 \documentclass[authoryear,preprint,review,12pt]{elsarticle} %THIS is the one used when submitted to journal

\usepackage{amssymb,amsthm,graphics,graphicx,epsfig,amsmath}
\usepackage{lineno}

\journal{Earth and Planetary Science Letters}

\begin{document}

\begin{frontmatter}

%% Title, authors and addresses

%% use the tnoteref command within \title for footnotes;
%% use the tnotetext command for the associated footnote;
%% use the fnref command within \author or \address for footnotes;
%% use the fntext command for the associated footnote;
%% use the corref command within \author for corresponding author footnotes;
%% use the cortext command for the associated footnote;
%% use the ead command for the email address,
%% and the form \ead[url] for the home page:
%%
%% \title{Title\tnoteref{label1}}
%% \tnotetext[label1]{}
%% \author{Name\corref{cor1}\fnref{label2}}
%% \ead{email address}
%% \ead[url]{home page}
%% \fntext[label2]{}
%% \cortext[cor1]{}
%% \address{Address\fnref{label3}}
%% \fntext[label3]{}

\title{Thermoelastic Properties of Ringwoodite (Fe$_x$,Mg$_{1-x}$)$_2$SiO$_4$: Its Relationship to the 520 km Seismic Discontinuity.}

%% use optional labels to link authors explicitly to addresses:
%% \author[label1,label2]{<author name>}
%% \address[label1]{<address>}
%% \address[label2]{<address>}
%%%%%%%%%%%%%%%%%%%%%
\author[address1]{Maribel N\'u\~nez Valdez\corref{cor1}}
\ead{valdez@physics.umn.edu}
\author[address2,address3]{Zhongqing Wu}
\author[address3]{Yonggang G. Yu}
\author[address4]{Justin Revenaugh}
\author[address3,address5]{Renata M. Wentzcovitch}

\address[address1]{School of Physics and Astronomy, University of Minnesota, Minneapolis, MN 55414, USA}
\address[address2]{School of Earth and Space Sciences, University of Science and Technology of China, Hefei, Anhui 230026, China}
\address[address3]{Department of Chemical Engineering and Materials Science, University of Minnesota, Minneapolis, MN 55414, USA}
\address[address4]{Department of Earth Sciences, University of Minnesota, Minneapolis, MN 55414, USA}
\address[address5]{Minnesota Supercomputing Institute, University of Minnesota, Minneapolis, MN 55414, USA}
\cortext[cor1]{Corresponding author. Tel.: +1 612 624 2872; fax: +1 612 626 7246}
%%%%%%%%%%%%%%%%%%%%%
\begin{abstract}
%% Text of abstract
We combine density functional theory (DFT) within the local density approximation (LDA), the quasiharmonic approximation (QHA), and a model vibrational density of states (VDoS) to calculate elastic moduli and sound velocities of $\gamma-$(Fe$_x$,Mg$_{1-x}$)$_2$SiO$_4$ (ringwoodite), the most abundant mineral of the lower Earth's transition zone (TZ). Comparison with experimental values at room-temperature  and high pressure or ambient-pressure and high temperature shows good agreement with our first-principles findings. Then, we investigate the contrasts associated with the  $\beta\rightarrow\gamma$(Fe$_x$,Mg$_{1-x}$)$_2$SiO$_4$ transformation at pressures and temperatures relevant to the TZ. This information offers clearly defined reference values to advance the understanding of the nature of the 520 km seismic discontinuity.  
\end{abstract}
%%%%%%%%%%%%%%%%%%%%%
\begin{keyword}
first principles \sep ringwoodite \sep elasticity \sep transition zone \sep 520 km discontinuity
%% keywords here, in the form: keyword \sep keyword
%% MSC codes here, in the form: \MSC code \sep code
%% or \MSC[2008] code \sep code (2000 is the default)
\end{keyword}

\end{frontmatter}

\linenumbers

%% main text

\section{Introduction}
Wadsleyite ($\beta$-phase) and ringwoodite ($\gamma$-phase) are the high-pressure polymorphs of olivine ($\alpha$-phase), (Fe$_x$,Mg$_{1-x}$)$_2$SiO$_4$. These minerals are the main constituents of the Earth's upper mantle (UM) \citep{Rinwood,Putnis} and transition zone (TZ) \citep{IR}. Under pressure, the transformation from olivine to wadsleyite happens at $\sim$13.5 GPa and from wadsleyite to ringwoodite at $\sim$18 GPa near 1600 K \citep{KI89,Ak89}. These transformations  are associated with two major discontinuities in seismic velocities in the Earth's interior at about 410 km and 520 km depth, respectively \citep{RJ}.  While the first discontinuity is a well characterized and sharp feature in seismic data, the second varies considerably with location. 

Experimental and computational approaches have been used to study properties of these minerals at {\it in situ} conditions and their relationship with seismic discontinuities.  Mg-end member and Fe-bearing $\alpha$-phases have been widely investigated at simultaneous high pressure and temperature. Computational results and experimental data on elastic properties and sound velocities seem to be consistent  with seismic measurements \citep[see e.g.][and references there in]{Nunez2010,Stack10,Li07}. On the other hand, even though great efforts have been made to obtain measurements of elastic properties and wave velocities of Fe-free and Fe-bearing wadsleyite and ringwoodite under high temperature and pressure using either ultrasonic or Brillouin scattering techniques \citep{Li96,Zha97,Isaak07,Sinog98,Li00,Liu09,Maya04,Isaak10,Li03,Higo06,Weidner84,Sinog03,Jackson00,Maya05}, results are still limited. Therefore large extrapolations from room conditions to conditions of the TZ are often used.   

First principles calculations employing the quasiharmonic approximation (QHA), valid up to about two thirds of the melting temperature, or molecular dynamics (MD) methods, valid near and above melting temperatures, complement each other and are used to obtain elastic moduli under high-pressure-temperature conditions.
Calculations of elastic constants using the QHA, though computationally less demanding than MD,  still required calculations of vibrational density of states (VDoS) for each strained atomic configuration at several pressures, that is, about 1000 parallel jobs \citep{PdS,DS}. 

In this paper we use the analytical and computational approach  by \cite{Wu2011} and tested on periclase-MgO, $\alpha-$Mg$_2$SiO$_4$,  and more recently on $\beta-$(Fe$_x$,Mg$_{1-x}$)$_2$SiO$_4$ \citep{MNVGRL2012}, to calculate bulk ($K$) and shear ($G$) moduli and sound velocities of the $\gamma-$(Fe$_x$,Mg$_{1-x}$)$_2$SiO$_4$ phase. This method uses only static elastic constants and phonon density of states for unstrained configurations, therefore reducing the amount of computational time and resources by one to two orders of magnitude. We then address contrasts across the $\beta\rightarrow\gamma$(Fe$_x$,Mg$_{1-x}$)$_2$SiO$_4$ transition near conditions of the 520 km seismic discontinuity.

%%%%%%%%%%%%%%%%%%%%%%%%%%
\section{Methodology}
\subsection{Computational Details}
Calculations based on Density Functional Theory (DFT) \citep{HK,KS} were performed using the local density approximation (LDA) \citep{Cep1980}. Ultrasoft pseudopotentials generated by the Vanderbilt method \citep{Vanderbilt} were used to describe Fe, Si, and O. A norm-conserving pseudopotential generated by the von Car method was used for Mg.  Further details about these pseudopotentials are given in \citep{Nunez2011}. Equilibrium structures of ringwoodite (28 atoms/cell) at arbitrary pressures were found using the variable cell-shape damped molecular dynamics approach  \citep{Wentz1991, Wentz1993} as implemented in the quantum-ESPRESSO (QE) code \citep{Giann}. The plane-wave kinetic energy cutoff used was 40 Ry and for the charge density 160 Ry. The {\bf k}-point sampling of the charge density was determined on a $2\times2\times2$ Monkhorst-Pack grid of the Brillouin Zone (BZ) shifted by $\left(\frac{1}{2} \frac{1}{2} \frac{1}{2}\right)$ from the origin. These parameters correspond to having interatomic forces smaller than 10$^{-4}$ Ry/a.u. and pressure convergence within 0.5 GPa. Dynamical matrices were obtained using density functional perturbation theory (DFPT) \citep{Baroni2001} via QE. At each pressure, a dynamical matrix was calculated on a $2\times2\times2$ {\bf q}-point mesh for one atomic configuration only. In principle about 10 other configurations should be used as well, but here we are more interested in frequencies with strain and the current approximation seems to be sufficiently accurate. Force constants were extracted and interpolated to a $12\times12\times12$ regular {\bf q}-point mesh to produce VDoS. 
%%%%%%%%%%%%%%%%%%%%%%%%%%
\subsection{High-Temperature-Pressure Elastic Theory}
Exploiting the information about the strain and volume dependence of phonon frequencies, we determine the thermal contribution to the Helmholtz free energy $F$ within the QHA \citep{Wallace}, that is,
\begin{eqnarray}
F\left(e,V,T\right)=U_{st}(e,V)+\frac{1}{2}\sum_{{\bf q},m}{\hbar\omega_{{\bf q},m}(e,V)}+\nonumber\\
+k_BT\sum_{{\bf q},m}\ln\left\{1-exp\left[-\frac{\hbar\omega_{{\bf q},m}(e,V)}{k_BT}\right]\right\},
\end{eqnarray}
where ${\bf q}$ is the phonon wave vector, $m$ is the normal mode index,  $T$ is temperature, $U_{st}$ is the static internal energy at equilibrium volume $V$ under isotropic pressure $P$ and infinitesimal strain $e$, $\hbar$ and $k_B$ are Planck and Boltzmann constants, respectively. Isothermal elastic constants are given by 
\begin{eqnarray}
C_{ijkl}^T=\left[\frac{\partial^2G(P,T)}{\partial e_{ij}e_{kl}}\right]_P, \label{cijth}
\end{eqnarray}
with $G=F+PV$, the Gibbs energy, and $i,j,k,l=1,\dots,3$. To convert to adiabatic elastic constants, one uses the relationship:
\begin{equation}
C_{ijkl}^S=C_{ijkl}^T+\frac{T}{VC_V}\frac{\partial S}{\partial e_{ij}}\frac{\partial S}{\partial e_{kl}}\delta_{ij}\delta_{kl}, 
\end{equation} 
where $C_V$ is heat capacity at constant volume, and $S$ is entropy. For orthorhombic crystals, the non-shear elastic constants  of Eq. (\ref{cijth}) are:
\begin{eqnarray}
C_{iijj}^T&=&\left[\frac{\partial^2F({\bf e},V,T)}{\partial e_{ii}e_{jj}}\right]_P+(1-\delta_{ij})P(V,T),\nonumber \\
&=&C_{iijj}^{st}(V)+C_{iijj}^{ph}(V,T),
\end{eqnarray}
while the shear elastic constants are:
\begin{equation}
C_{ijij}^T=C_{ijij}^{st}(V)+C_{ijij}^{ph}(V,T).
\end{equation}
Elastic constants $C_{iijj}^{ph}$ and $C_{ijij}^{ph}$ can be expressed as functions of the volume Gr\"uneisen parameters, $\gamma_{{\bf q},m}=-\partial(\ln \omega_{{\bf q},m})/\partial(\ln V)$:
\begin{equation}
\frac{d\omega_{{\bf q},m}}{\omega_{{\bf q},m}}=-\gamma_{{\bf q},m}\frac{d V}{V},
\end{equation}
or the generalization to strain  Gr\"uneisen parameters:
\begin{equation}
\frac{\partial\omega_{{\bf q},m}}{\omega_{{\bf q},m}}=-\gamma_{{\bf q},m}^{ij}e_{ij}.
\end{equation} 

We have used the Wu-Wentzcovitch method \citep{Wu2011} to compute the thermal contribution to the elastic constants, $C_{ijij}^{ph}(V,T)$. 
This method allows the computation of thermal elastic constants without performing phonon calculations for strained configurations with the approximation that strain and mode Gr\"uneisen parameters have isotropic distribution, which is equivalent to assuming that thermal pressure is isotropic. This is a good approximation  \citep{Carrier2007} implicit in the QHA calculation of thermal pressures. 

After obtaining VDoS at several volumes by first principles, average strain Gr\"uneisen parameters were computed at such volumes and interpolated in a fine volume-temperature grid, that was then inverted to a pressure-temperature grid of 0.1 GPa and 10 K spacings. Static elastic constants previously computed \citep{Nunez2011} were also used. 

Voigt and Reuss bounds of bulk and shear moduli of orthorhombic crystals at high temperature were calculated using adiabatic elastic constants as \citep{Watt76,Watt}:
{\setlength\arraycolsep{.5pt}
\small
\begin{eqnarray}
K_V&=&\frac{1}{9}\left[C_{11}+C_{22}+C_{33}+2\left(C_{12}+C_{13}+C_{23}\right)\right],\\
G_V&=&\frac{1}{15}\left[C_{11}+C_{22}+C_{33}+3\left(C_{44}+C_{55}+C_{66}\right)\right.\nonumber\\
&&\left.-\left(C_{12}+C_{13}+C_{23}\right)\right],\\
K_R&=&D\left[C_{11}\left(C_{22}+C_{33}-2C_{23}\right)+C_{22}\left(C_{33}-2C_{13}\right)\right.\nonumber\\
&&-2C_{33}C_{12}+C_{12}\left(2C_{23}-C_{12}\right)\nonumber\\
&&\left.+C_{13}\left(2C_{12}-C_{13}\right)+C_{23}\left(2C_{13}-C_{23}\right)\right]^{-1},\\
G_R&=&15\left\{4\left[C_{11}\left(C_{22}+C_{33}+C_{23}\right)+C_{22}\left(C_{33}+C_{13}\right)\right.\right.\nonumber\\
&&+C_{33}C_{12}-C_{12}\left(C_{12}+C_{23}\right)-C_{13}\left(C_{13}+C_{12}\right)\nonumber\\
&&\left.-C_{23}\left(C_{23}+C_{13}\right)\right]/D\nonumber\\
&&\left.+3\left(1/C_{44}+1/C_{55}+1/C_{66}\right)\right\}^{-1},\\
D&=&C_{13}\left(C_{12}C_{23}-C_{13}C_{22}\right)+C_{23}\left(C_{12}C_{13}-C_{23}C_{11}\right) \nonumber\\
&&+C_{33}\left(C_{11}C_{22}-C_{12}^2\right).
\end{eqnarray}}
With Voight-Reuss-Hill averages of elastic moduli, i.e.,
\begin{equation}
K=\frac{K_V+K_R}{2}\,\,\,\mbox{and}\,\,\,G=\frac{G_V+G_R}{2},
\end{equation}
isotropic sound velocities are given by:
\begin{equation}
\begin{aligned}
V_P=\sqrt{\frac{K+\frac{4}{3}G}{\rho}},\;\;\;\;V_S=\sqrt{\frac{G}{\rho}},\;\;\;\;V_\Phi=\sqrt{\frac{K}{\rho}}, \label{eqvels}
\end{aligned}
\end{equation}
where $\rho$ is density, and $V_P$, $V_S$, and $V_\Phi$ are compressional, shear and bulk velocities, respectively. 
%%%%%%%%%%%%%%%%%%%%%%%%
\begin{figure*}%[h]
\begin{center}
\includegraphics[width=140mm,height=130mm]{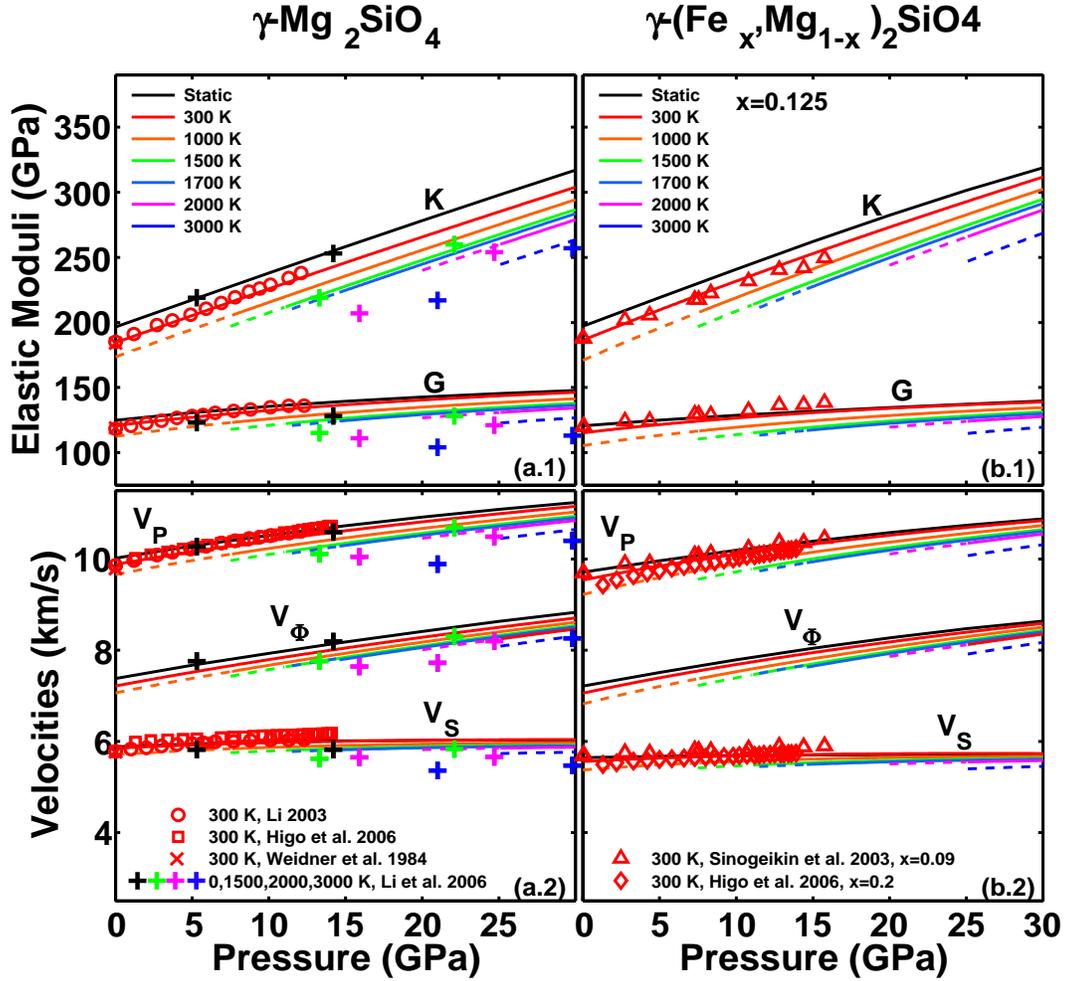}
\caption{(Color online) Pressure and temperature dependence of bulk modulus ($K$), shear modulus ($G$), compressional velocity ($V_P$), shear velocity ($V_S$) and bulk velocity ($V_\Phi$) for Fe-free ringwoodite (a,c) and Fe-bearing ringwoodite (b,d). First principles calculations within LDA (solid lines) are compared to available experimental data (symbols). Note, however, that low-pressure-high temperature calculated trends (dash lines) are outside the validity of the QHA.}\label{KGV_vs_P}
\end{center}
\end{figure*}
%%%%%%%%%%%%%%%%%%%%%%%%
%%%%%%%%%%%%%%%%%%%%%%%%%%
\section{Results}
\begin{figure*}%[h]
\begin{center}
\includegraphics[width=140mm,height=130mm]{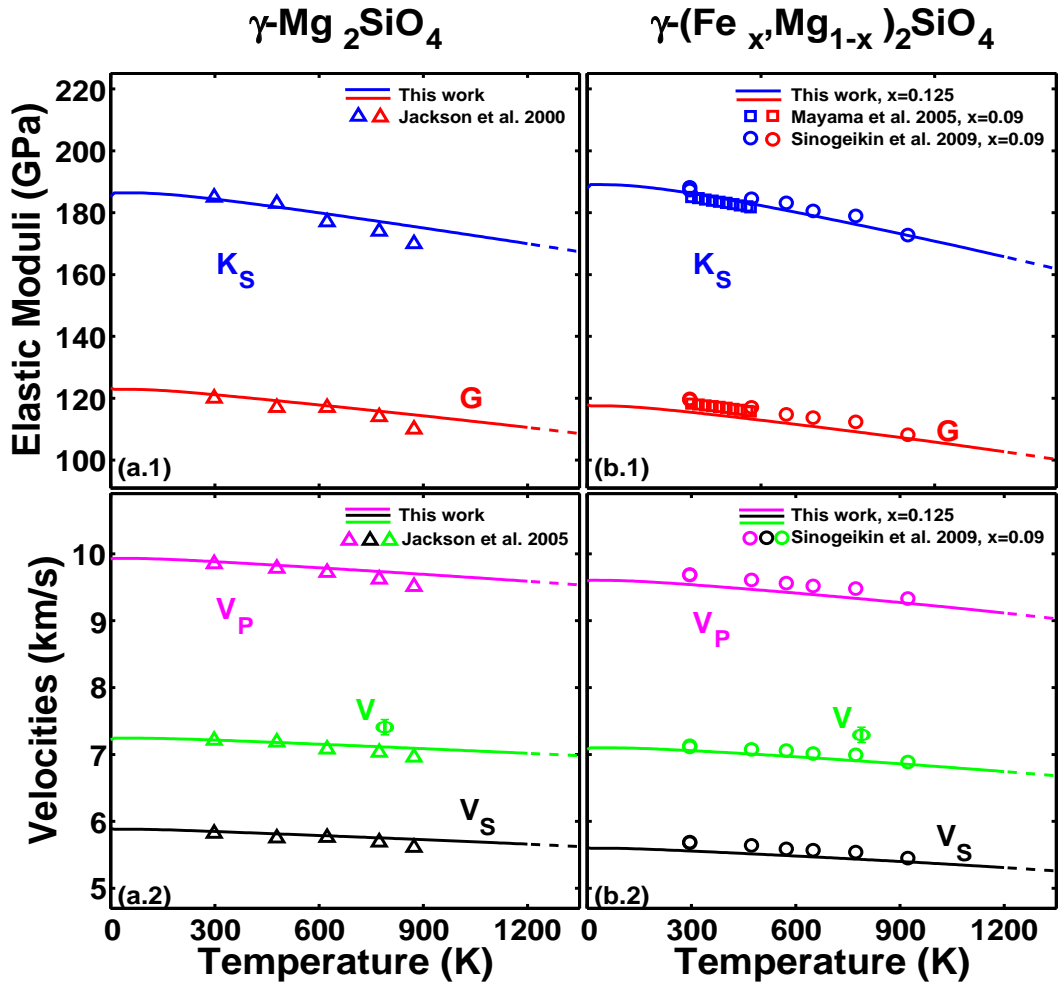}
\caption{(Color online) Temperature dependence of bulk modulus ($K$), shear modulus ($G$), compressional velocity ($V_P$), shear velocity ($V_S$) and bulk velocity ($V_\Phi$) for Mg-end member ringwoodite (a,c) and Fe-bearing ringwoodite (b,d). First principles calculations within LDA (solid lines) are compared to available experimental data (symbols) at P=0 GPa.}\label{KGV_vs_T}
\end{center}
\end{figure*}

We present first-principles results of aggregate properties of Fe-bearing ringwoodite at pressures and temperatures relevant to the TZ. All the approximations described in the previous section provided an excellent description of bulk and shear moduli, and sound velocities within the valid regime of the QHA established for the Fe-free quantities \citep{Yu08} for $\gamma-$(Fe$_x$,Mg$_{1-x}$)$_2$SiO$_4$ with $x=0$ and  $x=0.125$ (see Figs. \ref{KGV_vs_P} and \ref{KGV_vs_T}). 
%%%%%%%%%%%%%%%%%%%%%%%%
%%%%%%%%%%%%%%%%%%%%%%%%

In the case of the Mg-end member ringwoodite, $K$ increases more rapidly than $G$, as a function of pressure, and decreases faster than $G$ with increasing temperature. At 300 K the agreement between experimental data \citep{Li03,Higo06,Weidner84} and our DFT-results is truly excellent for elastic moduli (Fig. \ref{KGV_vs_P}a), and sound velocities (Fig. \ref{KGV_vs_P}c). Results from a molecular dynamics study by \cite{Li06} also compare well with our results for $K$, $V_P$, and $V_\Phi$ within the QHA limits. Our predicted $G$ and $V_S$ are larger  and smaller, respectively, than molecular dynamics values \cite{Li06}. This difference is primarily caused by the use of the GGA approximation in the MD simulation. Nevertheless the general agreement is good and it is the only other source to compare aggregate properties of $\gamma-$Mg$_2$SiO$_{4}$ at high pressures and temperatures.  Results of Fe-bearing ringwoodite as function of pressure are shown in Figs. \ref{KGV_vs_P}b and \ref{KGV_vs_P}d. Experimental data by \cite{Sinog03} at room temperature and $x=0.09$ are in excellent agreement with our 300 K curve, though after 10 GPa our $K$ tends to be larger, while $G$ tends to be smaller (Fig. \ref{KGV_vs_P}b). Fe-bearing compressional, shear and bulk velocities are smaller than their Fe-free counterparts (Fig. \ref{KGV_vs_P}d). Predictions for $x=0.125$ fall in between two experimental reports with iron concentrations of $x=0.09$ \citep{Sinog03} and $x=0.2$ \citep{Higo06}. From Fig. \ref{KGV_vs_P}d one can see that $V_P$ is the most affected by iron concentration and temperature, while $V_S$ seems to be the least affected by these two factors.   
%%%%%%%%%%%%%%%%%%%%%%%%
\begin{figure}%[h]
\begin{center}
\includegraphics[width=80mm,height=135mm]{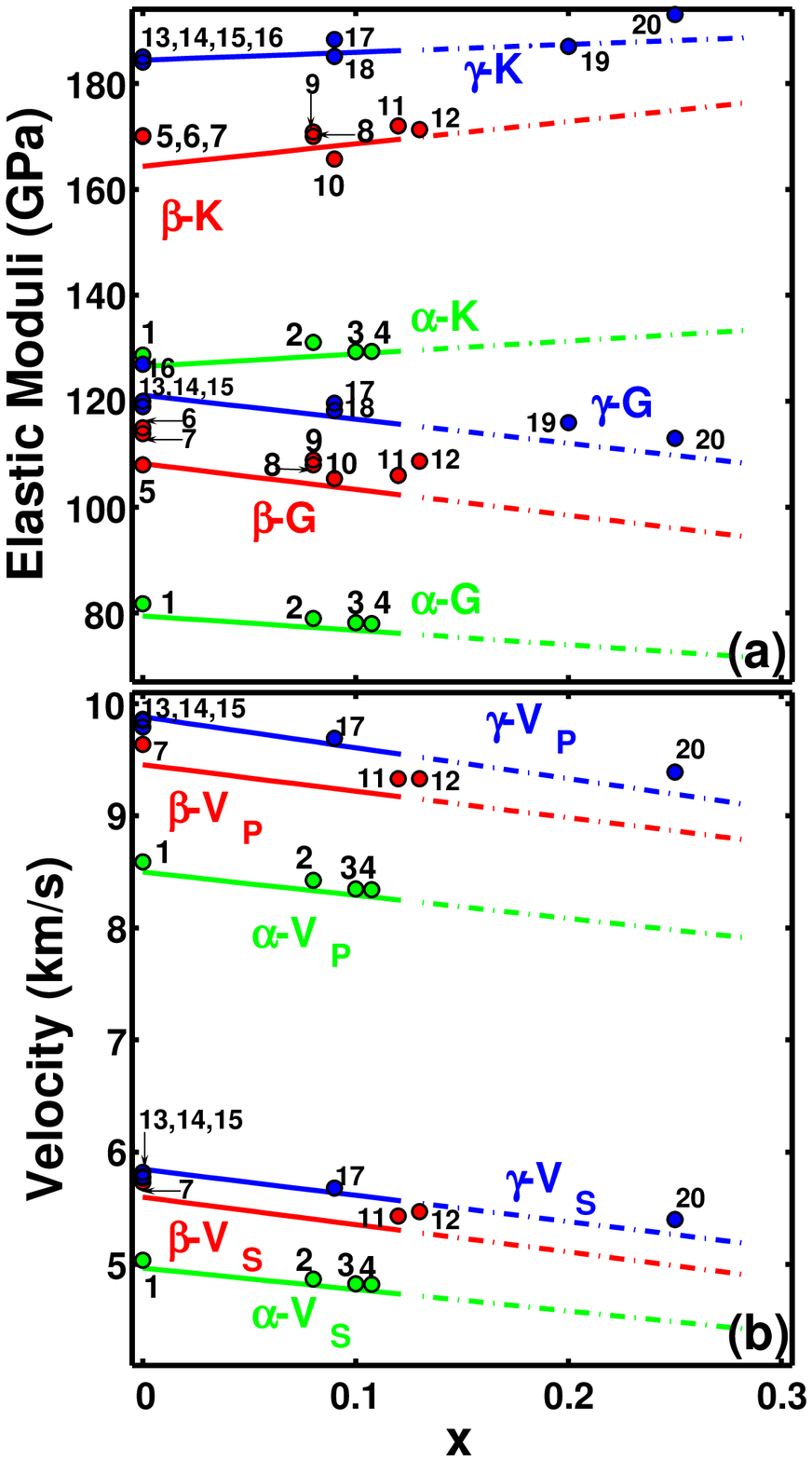}
\caption{(Color online) Dependence on low iron content at $P=0$ GPa and $T=300$ K of a) elastic moduli, $K$ and $G$,  and b) velocities, $V_P$ and $V_S$, (lines) compared to experimental data (circles): 1-\cite{Isaak89}; 2,3-\cite{Isaak92}; 4-\cite{Abramson}; 5-\cite{Li96}; 6-\cite{Zha97}; 7-\cite{Isaak07}; 8-\cite{Sinog98}; 9-\cite{Isaak10}; 10-\cite{Maya04}; 11-\cite{Li00}; 12-\cite{Liu09}; 13-\cite{Weidner84}; 14-\cite{Jackson00}; 15-\cite{Li03}; 16,19-\cite{Higo06}; 17-\cite{Sinog03}; 18-\cite{Maya05}; 20-\cite{Sinog97}.}\label{abg_KG_vs_x}
\end{center}
\end{figure}
%%%%%%%%%%%%%%%%%%%%%%%%

The temperature dependence of elastic moduli and sound velocities of Fe-free and Fe-bearing ringwoodite at ambient pressure are shown in Fig. \ref{KGV_vs_T}. For $x=0$, the agreement between experimental results \citep{Jackson00} and our findings is outstanding  (Figs. \ref{KGV_vs_T}a and \ref{KGV_vs_T}c). Similarly, our predicted aggregate properties for $x=0.125$ are in excellent correspondence with experiments having $x=0.09$ \citep{Maya05,Sinog03} (Figs. \ref{KGV_vs_T}b and \ref{KGV_vs_T}d). Predicted Fe-bearing $K$ is larger than Fe-free $K$ at 300 K (Table \ref{Table1}), and $dK/dT$ is more negative for $x=0.125$ than for $x=0$. On the other hand, $G$ decreases with iron, but likewise $K$, $dG/dT$ is more negative for $x=0.125$ than for $x=0$.
Predicted compressional, shear, and bulk velocities as function of temperature are smaller than those reported by \cite{Sinog03}, which can be understo0d given the difference in iron content (Fig. \ref{KGV_vs_T}d).
%%%%%%%%%%%%%%%%%%%%%%%%
\begin{figure}%[h]
\begin{center}
\includegraphics[width=80mm,height=170mm]{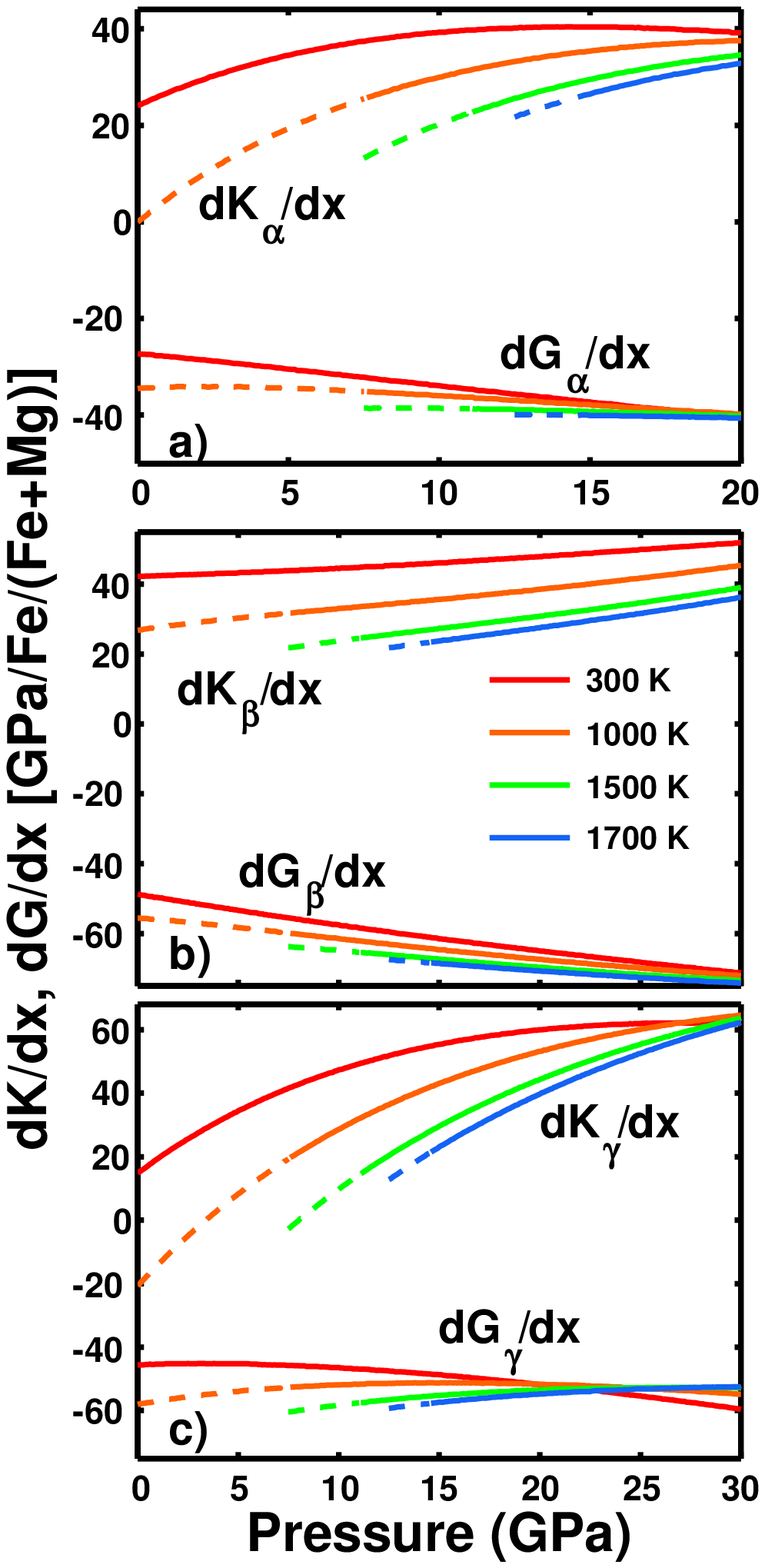}
\caption{(Color online) Pressure dependence of $dK/dx$ and $dG/dx$ for a) olivine, b) wadsleyite, and c) ringwoodite.}\label{dKdGdx_vs_P}
\end{center}
\end{figure}
%%%%%%%%%%%%%%%%%%%%%%%%
The $x$ dependence of elastic moduli and velocities for $\alpha-$, $\beta-$, and $\gamma-$(Fe$_x$,Mg$_{1-x}$)$_2$SiO$_4$   at high temperatures and pressures is shown in Fig. \ref{abg_KG_vs_x}. We find $dK/dx$ to be positive for all three phases, while $dG/dx$, $dV_P/dx$, $dV_S/dx$, and $dV_\Phi/dx$ are negative. For small $x$, a linear trend given by our results of elastic moduli and velocities compares well to experimental data of olivine and ringwoodite. On the other hand,  experimental values of wadsleyite are more scattered and deviate the most from the proposed linear behavior. Detailed dependence on pressure and temperature of $dK/dx$, $dG/dx$, $dV_P/dx$, $dV_S/dx$, and $dV_\Phi/dx$ are shown in Figs. \ref{dKdGdx_vs_P}  and \ref{dVdx_vs_P}. $dK/dx$ and $dG/dx$ for all three phases exhibit qualitatively similar behavior in the pressure range considered. At low pressure they are quite sensitive to temperatures, but they seem to converge at high pressure (Figs. \ref{dKdGdx_vs_P}a and \ref{dKdGdx_vs_P}c).  $dV_P/dx$, $dV_S/dx$, and $dV_\Phi/dx$ for all three phases are also more sensitive to temperature at lower pressures. 
%%%%%%%%%%%%%%%%%%%%%%%%
\begin{figure*}%[h]
\begin{center}
\includegraphics[width=140mm,height=55mm]{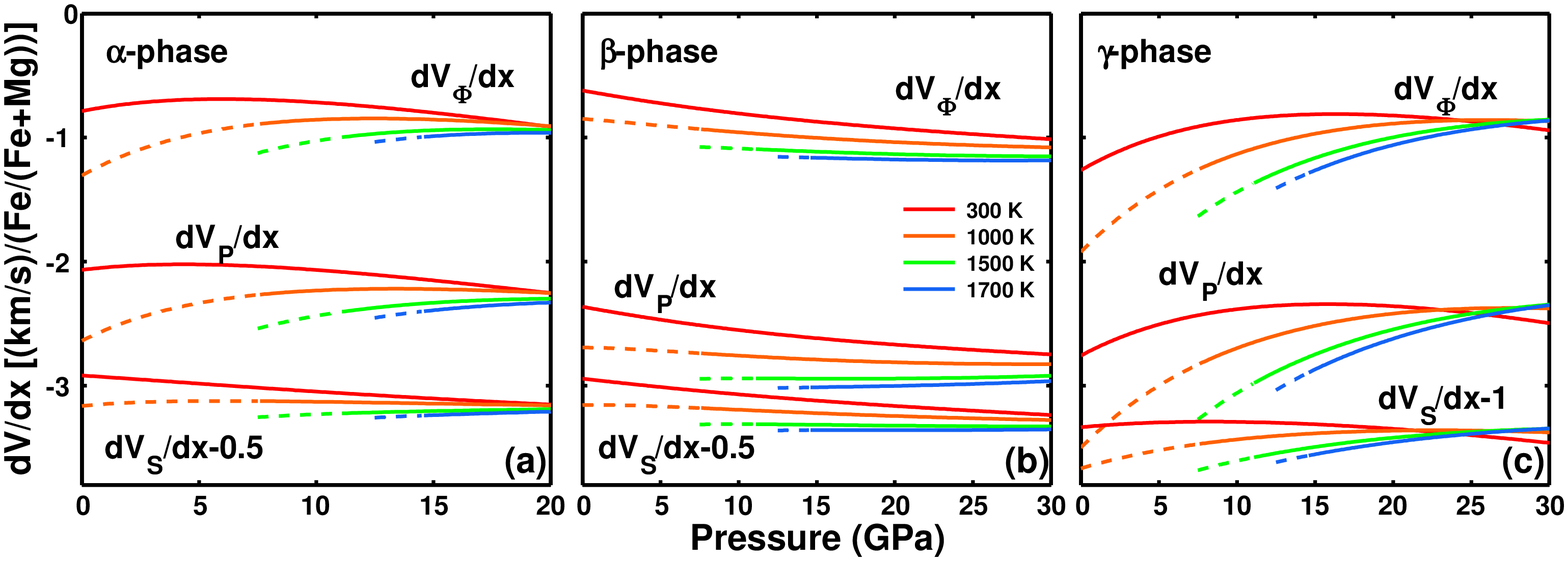}
\caption{(Color online) Pressure dependence of $dV/dx$  for a) olivine, b) wadsleyite, and c) ringwoodite.}\label{dVdx_vs_P}
\end{center}
\end{figure*}
%%%%%%%%%%%%%%%%%%%%%%%%

%%%%%%%%%%%%%%%%%%%%%%%%%%
\section{Geophysical implications: The 520 km Discontinuity}
The seismic discontinuity near 520 km depth is often attributed to the phase change of wadsleyite to ringwoodite \citep{KI89,Shearer90,RJ}.   It is more likely to be unobserved than either of its near neighbors at 410-km and 660-km depth.   It is, on average, a smaller amplitude feature \citep[e.g.][]{RJ} such that less frequent observation is to be expected.  In some studies the 520 km discontinuity appears as a split arrival or doublet \citep[e.g.][]{DW01,Cham05,Bag09}.  When split, the two discontinuities are observed at depths of approximately 500 and 560 km \citep{DW01}.  Notably the sum of the two seismic features is larger than typical non-split observations.  Whether this is the result of an upward bias in identifying split arrivals or the result of greater net velocity contrast it is not clear. Therefore accurate data on elasticity of wadsleyite and ringwoodite are critical for investigating the role of the transformation $\beta\rightarrow\gamma$ on the 520 km seismic discontinuity. 
%%%%%%%%%%%%%%%%%%%%%%%%
\begin{figure*}%[h]
\begin{center}
\includegraphics[width=140mm,height=130mm]{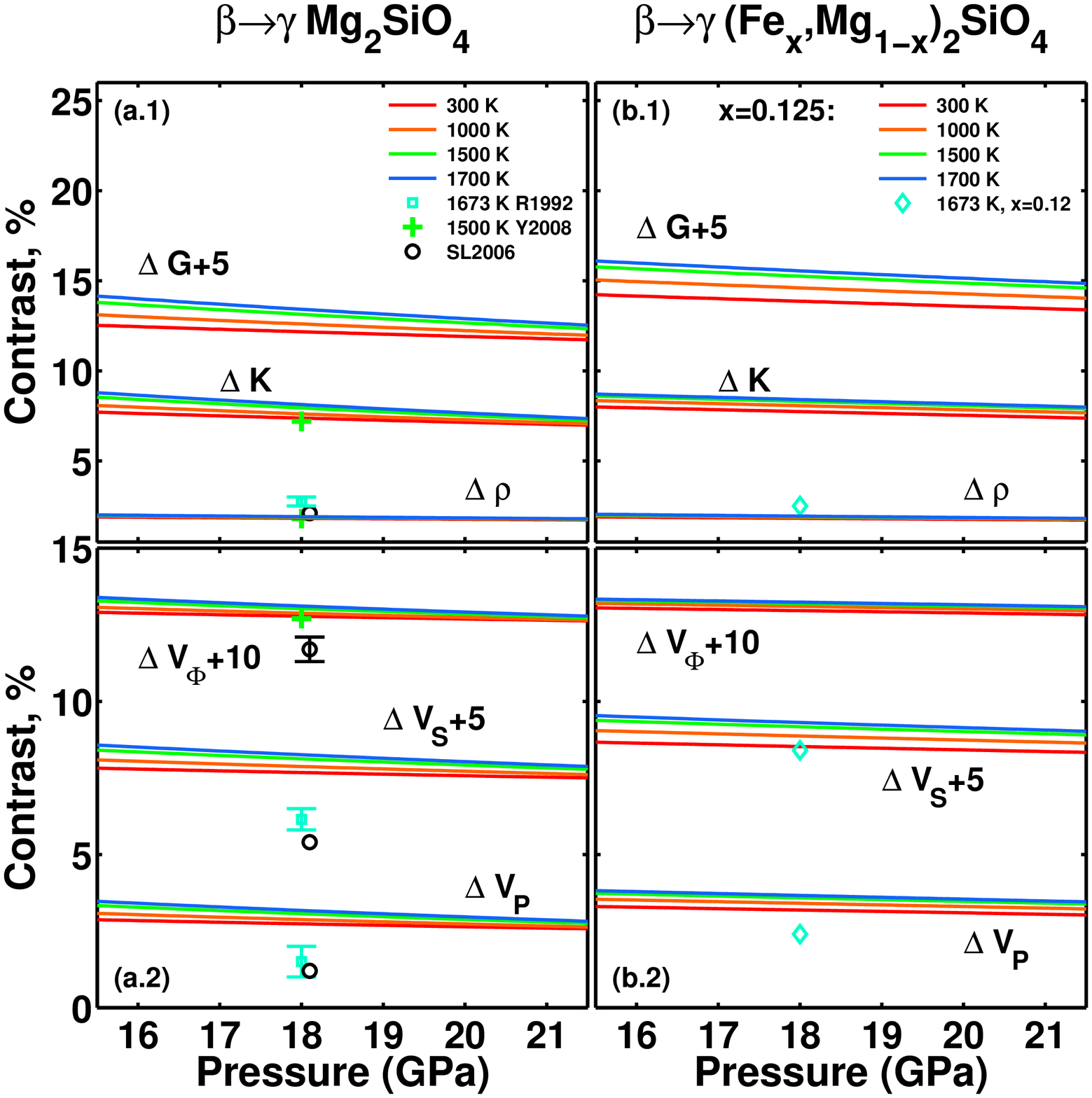}
\caption{(Color online) Density, elastic, and velocity contrasts (lines) compared to laboratory and seismic data across the Fe-free (a,c) and Fe-bearing (b,d) $\beta\rightarrow\gamma$ transition. R1992: \cite{Rigden92}; Y2008: \cite{Yu08}; SL2006: \cite{LS06}; $x=0.12$: \cite{Sinog03}.}\label{Fig5}
\end{center}
\end{figure*}
%%%%%%%%%%%%%%%%%%%%%%%%
We use our results on aggregate properties of $\beta-$ \citep{MNVGRL2012} and $\gamma-$(Fe$_x$,Mg$_{1-x}$)$_2$SiO$_4$ at temperatures and pressures encompassing the TZ to estimate the magnitude of the discontinuity across the phase transition. Although this is a divariant phase transition and calculation of the two-phase loop is beyond the scope of this work, we can clearly calculate velocity increases throughout the entire transition.  As we saw in the previous section, experimental studies dealing with simultaneous high pressures and temperatures offer limited data, and analyses usually extrapolate results at ambient conditions either in temperature or pressure to conditions near 520 km depth ($\sim$18 GPa and $\sim$1600 K). The lack of other source of knowledge makes it difficult to outline conclusions and/or explain the nature of the 520 km seismic discontinuity. With this paper we hope to advance the understanding of this discontinuity. 

To quantify the magnitude of the discontinuity across the $\beta$ to $\gamma$ transition at finite temperatures we use the contrast $\Delta$ of a particular property $M$ defined as:
\begin{equation}  
 \Delta M=\frac{\left(M_{x,\gamma}- M_{x,\beta}\right)}{\frac{\left(M_{x,\beta}+ M_{x,\gamma}\right)}{2}}\times 100,
\end{equation} 
where $M$ could be density, elastic modulus, or velocity. Table \ref{Table2} and Fig. \ref{Fig5} show our calculated contrasts at finite temperatures for the $\beta\rightarrow\gamma$ transition in Fe-free and Fe-bearing phases. We first notice that $\Delta \rho$ is almost independent of temperature and pressure with iron having an insignificant effect (Figs. \ref{Fig5}a and \ref{Fig5}b), as suggested by \cite{Yu08}. Extrapolated experimental results and seismic data also suggested that iron should have a small effect on this quantity \citep{Rigden91,LS06,Sinog03}. For $x=0$ we obtain a very good agreement with a previous estimation by \cite{Yu08}. For $x=0.125$ our result is slightly smaller than that by \cite{Sinog03}  for $x=0.09$. 
Contrasts of elastic moduli (Figs. \ref{Fig5}a and \ref{Fig5}b) indicate that $\Delta G$ is more sensitive to temperature than $\Delta K$. This dependence is greater in $x=0.125$ than in $x=0$.  In the 16--21GPa range, both $\Delta G$ and $\Delta K$ slightly decrease with pressure,  irrespective of $x$. Figs. \ref{Fig5}c and \ref{Fig5}d show that velocity contrasts, $\Delta V_P$, $\Delta V_S$, and $\Delta V_\Phi$, exhibit weak pressure dependence, decreasing  only slightly with pressure. Temperature affects $\Delta V_S$ more than $\Delta V_P$ or $\Delta V_\Phi$. For $x=0$, a comparison of our results to extrapolations of experimental data by \cite{Rigden91} to 18 GPa and 1673 K  and seismic data by \cite{LS06} shows predicted contrasts to be larger. $\Delta V_\Phi$ is in better agreement with a previous prediction at 1700 K \citep{Yu08} and with a contrast from seismic studies \citep{LS06}. Velocity contrasts increase with increasing $x$, however $\Delta V_S$ is the most enhanced by $x$, followed by $\Delta V_P$, and $\Delta V_\Phi$. Extrapolations of  $\Delta V_P$ and $\Delta V_S$ to 18 GPa and 1673 K with $x=0.12$ reported by \cite{Sinog03} are smaller than our predicted contrasts, though $\Delta V_S$ is closer to our prediction. 

%%%%%%%%%%%%%%%%%%%%%%%%
\begin{table*}
\scriptsize{
\centering          
\begin{tabular}{l  l  l  l  l }
\hline\hline
\multicolumn{5}{c}{$\beta-$(Mg$_{1-x}$,Fe$_{x}$)$_2$SiO$_{4}$}\\
\hline
$V$ (\AA${}^3$) & $K$ (GPa) & $G$  (GPa) & $x$ &Reference\\
\hline
541.3 & 164.4 & 107.7 & 0 & \cite{MNVGRL2012}, Single crystal/DFT \\
- &170 & 108 & 0 & \cite{Li96}, Poly-crystal/US \\
535.8(0.2) &170(2) & 115(2) & 0 & \cite{Zha97}, Single crystal/BS\\
- & 169.7(1.9)-170.7(2) & 113.9(0.7)-114.1(0.8) & 0 & \cite{Isaak07}, Poly-crystal/RUS \\
541.5 & 169.5 & 101.7 & 0.125 & This study, Single crystal/DFT \\
539.4(4) & 170(3) & 108(2) & 0.08 & \cite{Sinog98}, Single crystal/BS \\
- & 170.8(1.2) & 108.9(0.4) & 0.08 & \cite{Isaak10}, Poly-crystal/RUS\\
- & 165.72(6) & 105.43(2) & 0.09 & \cite{Maya04}, Poly-crystal/RUS \\
- & 172(2) & 106(1) & 0.12 & \cite{Li00}, Poly-crystal/US\\
- & 171.3(3) & 108.7(2) & 0.13 & \cite{Liu09}, Poly-crystal/US\\
\hline \hline
\multicolumn{5}{c}{$\gamma-$(Mg$_{1-x}$,Fe$_{x}$)$_2$SiO$_{4}$}\\
\hline

$V$ (\AA${}^3$) & $K$ (GPa) & $G$  (GPa) & $x$ &Reference\\
\hline
527.6 & 184.4 & 120.95 & 0 &  This study, Single crystal/DFT \\
525.3 & 184 & 119 & 0 & \cite{Weidner84}, Single crystal/BS\\
525.3 & 185(3) & 120.4(2) & 0 & \cite{Jackson00}, Single crystal/BS\\
- & 185(2) & 120(1) & 0 & \cite{Li03}, Poly-crystal/US\\
- & 185(2) & 127(1) & 0 & \cite{Higo06}, Poly-crystal/US\\
527.7 & 186.3 & 115.3 & 0.125 & This study, Single crystal/DFT \\
526.2(4)&188.3(30) & 119.6(20) & 0.09 & \cite{Sinog03}, Single crystal/BS\\
- & 185.11(0.16)-185.17(0.17) & 118.27(0.06) & 0.09 & \cite{Maya05}, Poly-crystal/RUS \\
- & 187(2) & 116(1) & 0.20 &\cite{Higo06}, Poly-crystal/US \\
\hline
\hline
\end{tabular}
\caption{Results on wadsleyite and ringwoodite for volume ($V$), bulk ($K$), and shear ($G$), moduli at ambient pressure and temperature. US: Ultrasonic techniques; BS: Brillouin scattering techniques; RUS: Resonant Ultrasonic techniques.}\label{Table1}
}
\end{table*}
%%%%%%%%%%%%%%%%
In an attempt to explain the presence of two mid-transition zone discontinuities, \cite{Saikia08} investigated the   solubility of CaSiO$_3$ perovskite  in majorite garnet at high-pressure (15--24 GPa) and  high-temperature   (1400$^\circ$ C--1600$^\circ$ C). They concluded that  in fertile peridotite   (i.e., peridotite enriched in  Ca  and  Al)   at   1400$^\circ$ C,   the  wadsleyite to ringwoodite phase  change produces  a   strong  discontinuity   at  $\sim$500--520 km depth,  while  the   exsolution of Ca-perovskite   produces   a   weak    discontinuity   near   540   km.  At   1600$^\circ$ C,   the   two   merge   to   form  a   single   discontinuity   at  540--560 km.  In   MORB-like compositions, wadsleyite    and  ringwoodite   are effectively absent,   but   exsolution  of   Ca-perovskite   causes   a  velocity discontinuity    near   560 km.   \cite{Saikia08}   concluded   that   a mechanical mixture (or seismically averaged  assemblage) of peridotite and MORB  would   have   two  discontinuities:   one   near   500 km   due   to   the   wadsleyite    to   ringwoodite  phase   change,   and   a  second  near   560 km   due   to   the   exsolution   of   Ca-perovskite   from   garnet, consistent with a hypothesis of \cite{DW01}.  

Of some question here is the association of the MORB-like mantle with the non-MORB component.  If MORB is associated with peridotitic mantle, the strength of the 500 km discontinuity would decrease with respect to that of the 520 km proportionally to the percentage of MORB. Since $\Delta V_S$ is smaller for the ex-solution of Ca-perovskite from majorite garnet in MORB-like material by a factor of 2 \citep{Saikia08}, the summed strength of the two seismic discontinuities, at 500 km and 560 km, in a MORB plus peridotite mixture would be smaller than that of the 520 km alone. This is opposite to observations  \citep{DW01}, and whether this is the result of an upward bias in identifying split arrivals or the result of greater net velocity contrast is not clear. If MORB is associated with MORB-depleted mantle (harzburgite or dunite) the greater wadsleyite fraction in the depleted mantle might compensate and increase in $\Delta V_S$ for the 500 km discontinuity. However, the smaller fraction of iron in MORB-depleted mantle \citep{JG80} should decrease $\Delta V_S$ (as shown in Table \ref{Table2}) of the  $\beta$ to $\gamma$ transition. The extent to which these effects counter-act each other and whether the greater summed strength of the two discontinuities can be explained by this picture, remains an open question. 
%%%%%%%%%%%%%%%%
\begin{table}
\centering          
\begin{tabular}{ l | r | r | r | r}
\hline\hline
T (K) &\multicolumn{2}{c |} {1500} &\multicolumn{2}{c }{1700 }\\
\hline
$x$& 0.0&0.125&0.0&0.125\\
\hline
$\Delta \rho$ & 1.89 & 1.92 & 1.91 & 1.94\\
$\Delta K$ & 7.94 & 8.31 & 8.13 & 8.41\\
$\Delta G$ & 8.14  & 10.26 & 8.42 & 10.55\\
$\Delta V_P$ & 3.07 & 3.58 & 3.17 & 3.67\\
$\Delta V_S$ & 3.12 & 4.17 & 3.26 & 4.31\\
$\Delta V_\Phi$ & 3.03 & 3.20 & 3.11 & 3.23\\
$\Delta (\rho V_P)$ & 4.96 & 4.99 & 5.08 & 5.11 \\
$\Delta (\rho V_S)$ & 5.02 & 5.04 & 5.17 & 5.20\\
\hline\hline
\end{tabular}
\caption{Predicted contrasts in \% across the $\beta\rightarrow\gamma$-(Mg$_{1-x}$,Fe$_{x}$)$_2$SiO$_{4}$   transition at 18 GPa.}\label{Table2}
\end{table}
%%%%%%%%%%%%%%%%%%%%%%%%%%
\section{Conclusions}
For the first time, we have presented parameter-free first-principles results of high pressure and high temperature aggregate elastic properties and sound velocities of Fe-bearing  ringwoodite. We used the QHA and a novel method of calculating elasticity at high temperatures \citep{Wu2011}. Treatment of strain Gr\"uneisen parameters via isotropic averages reduced greatly the computational cost of the task, which, otherwise, would have been much more intensive and lengthy. Within the QHA limit, our predictions for elastic and acoustic properties for $x=0$ were found to be in very good agreement with available experimental data at 300 K and ambient pressure  \citep{Li96,Zha97,Isaak07,Li03,Higo06,Weidner84} and molecular dynamics simulations \citep{Li06} at high pressures and temperatures. For $x=0.125$ our results compared very well with experimental data in the range $x=0.08-0.2$ \citep{Sinog98,Li00,Liu09,Sinog03,Higo06}. High-temperature and ambient pressure results in ringwoodite also reproduced experimental trends very well \citep{Isaak07,Maya04,Isaak10,Jackson00,Maya05,Sinog03} for Fe-free and Fe-bearing samples.

Overall our predictions showed well defined changes in the elastic and acoustic properties of the $\beta-$ and $\gamma-$ phases near conditions of the 520 km seismic discontinuity.  We show that pressure tends to decrease contrasts across the  $\beta\rightarrow\gamma$ transition while temperature and iron concentration tend to enhance them. 
The absence of global observations of the 520 km discontinuity could suggest regions of the TZ with less iron and/or smaller olivine content and irregular temperature topography. Other considerations to try to explain the intermittent nature of the discontinuity would involve changes in the pyroxene/garnet/Ca-pv system and the amount of water present in the TZ. These issues will be addressed in future similar studies including other relevant phases.
%%%%%%%%%%%%%%%%%%%%%%%%%%
\section{Acknowledgments}
Research supported by the NSF/EAR-1019853, and EAR-0810272.
Computations were performed using the VLab cyberinfrastructure at the Minnesota Supercomputing Institute.

\bibliographystyle{elsarticle-harv}
%\bibliography{<your-bib-database>}

\newpage

\newpage

\end{document}